\documentclass[floatfix,aps,prd,showpacs,amsmath,nofootinbib]{revtex4}
\usepackage{amsmath,amssymb,epsfig}
\newcommand{\q}{\quad} \newcommand{\qq}{\quad\quad}
\newcommand{\dc}{\displaystyle}
\newcommand{\const}{\text{const}}
\begin{document}

\title{Weak-field limit of conformal Weyl gravity}
\author{O.V. Barabash}\email{obar@univ.kiev.ua}
\author{H.P. Pyatkovska}
\affiliation{Department of Physics, Shevchenko National University,
Kiev 252022, Ukraine}

\date{\today}

\begin{abstract}

We study the weak-field limit of the conformal Weyl gravity
suggested by Mannheim as an alternative to Einstein's General
Relativity modeling both dark matter and dark energy. We solve the
field equations of the theory in the weak-field approximation for an
arbitrary spherically symmetric static distribution of matter in the
physical gauge with constant scalar field. Analysing the obtained
solution, we conclude that the conformal theory of gravitaty is
inconsistent with the Solar-system observational data.

\end{abstract}

\maketitle
\large
\section{Introduction}
In the Weyl conformal theory the action for the gravitation and
matter fields (the scalar field $S$ and the fermion field  $\psi$)
reads \cite{7}
$$
I_W=\int\!d^4x\sqrt{-g}\Bigl(-\alpha_gC_{\lambda\mu\nu\sigma}C^{\lambda\mu\nu\sigma}-\partial^\mu S\partial_\mu
S/2-\lambda S^4+S^2R/12{}-{}i\bar\psi\gamma^\mu(x)\nabla_\mu\psi+\zeta S\bar\psi\psi\Bigr),
$$
where $C_{\lambda\mu\nu\sigma}$ is the Weyl tensor, and $\alpha_g$, $\lambda$
and $\zeta$ are dimensionless constants (in units $c=\hbar=1$). We use the
metric signature ($-$,+,+,+). This action is invariant under the conformal
transformation of the metric and fields:
$$g_{\mu\nu}\rightarrow\Omega^2g_{\mu\nu},\q\psi\rightarrow\Omega^{-3/2}\psi,\q S\rightarrow S/\Omega.$$
Variation of the action with respect to $g_{\mu\nu}$, $S$ and $\psi$ gives the
following equations of motion \cite{5,8}:
\begin{equation}\label{eq}
4\alpha_g W_{\mu\nu}=T_{\mu\nu},
\end{equation}
\begin{equation}\label{eq2}
S^\mu_{;\mu}+SR/6-4\lambda S^3+\zeta\bar\psi\psi=0,
\end{equation}
\begin{equation}\label{eq3}
i\gamma^\mu\nabla_\mu\psi-\zeta S\psi=0,
\end{equation}
where
\begin{eqnarray}\label{tei1}
T_{\mu\nu}= 2S_\mu S_\nu/3-g_{\mu\nu}S^\alpha
S_\alpha/6-SS_{\mu;\nu}/3+g_{\mu\nu}SS^\alpha_{;\alpha}/3-
S^2(R_{\mu\nu}-g_{\mu\nu}R/2)/6-\lambda S^4g_{\mu\nu} \nonumber \\
{} +i\bar\psi\gamma_\mu\nabla_\nu\psi+g_{\mu\nu}(\zeta
S\bar\psi\psi-i\bar\psi\gamma^\lambda\nabla_\lambda\psi).
\end{eqnarray}
The tensor $W_{\mu\nu}$ was obtained for the first time in \cite{5} and is
expressed via the Ricci tensor and its derivatives.

In the conformal theory, material bodies do not move along geodesics (the
geodesic equation is not conformally invariant), but rather along trajectories
described by the equation \cite{9}
\begin{equation}\label{19}
\frac{d^2x^\lambda}{ds^2}+\Gamma^\lambda_{\mu\nu}\frac{dx^\mu}{ds}\frac{dx^\nu}{ds}
=-\frac{S_\mu}{S}\biggl(g^{\lambda\mu}+\frac{dx^\lambda}{ds}\frac{dx^\mu}{ds}\biggr).
\end{equation}

Equation (\ref{19}) as well as equations (\ref{eq}), (\ref{eq2}) and
(\ref{eq3}) are conformally invariant. The conformal invariance of
these equations guarantees independence of the physical effects from
the particular choice of gauge.  However, it is convenient to fix
the gauge by the condition
$$S=S_0=\const.$$
In this gauge, we arrive at the standard theory of the massive fermion field
$\psi$ and the usual geodesic equation of motion for a free body, which is
especially important for our investigation. The geodesic equation is verified
in standard gravitational tests both for massive bodies and for light rays if
the Schwarzschild space-time metric is used.\footnote{There exist, however,
radar-tracking observations of the Pioneer 10/11 spacecraft and others,
detecting an additional anomalous acceleration of these bodies \cite{6}
directed towards the Sun with magnitude $\sim 8.5 \times
10^{-10}$~m~$\cdot$s$^{-2}$. An attempt to explain this additional acceleration
in the conformal theory is made in \cite{1}.} Thus, we have a good test for
checking the conformal theory: in the gauge $S=S_0=\const$, the exterior metric
for a massive source should reduce to the Schwarzschild metric on the
Solar-system scale.

It is easy to verify that the Schwarzschild metric
\begin{equation}\label{shw}
{}-g_{00}=g_{11}^{-1}=1-\frac{2m}{r}
\end{equation}
is a solution of equations (\ref{eq}) in vacuum ($\psi=0$) on small scales
where one can neglect the cosmological term $\lambda S_0^4g_{\mu\nu}$ in
(\ref{tei1}). Indeed, it is known that metric (\ref{shw}) is a solution of the
equation $R_{\mu\nu}=0$, and the last equation implies that the tensor
$W_{\mu\nu}$ also vanishes because it is constructed from the Ricci tensor and
its derivatives only. We also have $T_{\mu\nu}=0$ in the vacuum ($\psi=0$),
which follows from (\ref{tei1}) in the gauge $S=S_0$.

However, the given reasoning has only formal mathematical character and does
not correspond to the physical formulation of the problem. The correct metric
outside the source should match the interior metric. The matching conditions
may not be satisfied for the exterior solution (\ref{shw}) or lead to a
negative value of the constant $m$. Thus, for example, an exterior spherically
symmetric solution of equations (\ref{eq}) in the gauge $S=S_0$ in the
weak-field approximation was obtained in \cite{3}. For the metric of the form
\begin{equation}\label{metrik1} \begin{array}{l}
ds^2 = -B(r)dt^2+A(r)dr^2+r^2d\Omega_2 \, , \medskip \\
B(r) = 1-b(r)\, , \q A(r)=1+a(r)\, ,
\end{array}
\end{equation}
the solution reads \cite{3}
\begin{equation}\label{metrik3}
\begin{array}{l}
\dc a(r)=\frac{2m}{r}+N\left[\frac{\sin
(kr+\phi)}{r}-k\cos(kr+\phi)\right]\,  , \medskip \\
\dc b(r)=\frac{2m}{r}+2N\,\frac{\sin (kr+\phi)}{r}\, .
\end{array}
\end{equation}
Assuming that the constant $m>0$ and setting $\phi=0$, one can ensure that
metric (\ref{metrik3}) really reduces to the Schwarzschild metric at small
distances (where $kr\ll 1$). However, as we will see below [see (\ref{20})],
the matching conditions imply that both assumptions ($m>0$ and $\phi=0$) become
untrue!

\section{Solution of the field equations in matter}
The energy-momentum tensor (\ref{tei1}) in the gauge $S=S_0$ in the
hydrodynamic approximation is given by
\begin{equation}\label{tei2}
T_{\mu\nu}=\varepsilon u_\mu u_\nu-S_0^2(R_{\mu\nu}-g_{\mu\nu}R/2)/6-\lambda S_0^4g_{\mu\nu}.
\end{equation}
Note that the tensor $W_{\mu\nu}$ is identically traceless, which is a
consequence of the conformal invariance of the theory. Taking the trace of
$T_{\mu\nu}$ (\ref{tei2}) and equating it to zero (due to equation (\ref{eq})
and the property of the tensor $W_{\mu\nu}$ mentioned above), we find
\begin{equation}\label{1}
R=24\lambda S_0^2+\frac{6}{S_0^2}\varepsilon.
\end{equation}

For a spherically symmetric and static distribution of matter, the metric has
the standard form (\ref{metrik1}). Performing transformation of the radial
coordinate, it is convenient to write this metric in the form
\begin{equation}\label{metrik2}
ds^2=C^2(\rho)[-D(\rho)dt^2+d\rho^2/D(\rho)+\rho^2d\Omega_2],
\end{equation}
where $r(\rho)=\rho C(\rho)$, and the functions $C(\rho)=1+c(\rho)$ and
$D(\rho)=1-d(\rho)$ are expressed through the metric coefficients $A(r)=1+a(r)$
and $B(r)=1-b(r)$ by the relations (in the linear approximation)
\begin{equation}\label{55}
a(r)=d(r)-2rc'(r),\quad b(r)=d(r)-2c(r).
\end{equation}
It is necessary to consider the quantities $a(r)$, $b(r)$, $c(\rho)$ and
$d(\rho)$ as small ones, of the order of smallness $\sim\epsilon\ll 1$ (the
applicability condition of the linear approximation). Due to spherical symmetry
and stationarity, the first set of equations (\ref{eq}) will give two
independent equations which are conveniently chosen to be
\begin{equation}\label{22}
4\alpha_g (W^0_0-W^1_1)={\cal T}^0_0-{\cal T}^1_1,\quad 4\alpha_g W_{11}={\cal T}_{11},
\end{equation}
where
$$
{\cal T}_{\mu\nu}=\varepsilon u_\mu u_\nu+\varepsilon g_{\mu\nu}/4-S_0^2(R_{\mu\nu}-g_{\mu\nu}R/4)/6
$$
is the traceless part of the energy-momentum tensor $T_{\mu\nu}$, and the index
``{\small 1}'' labels the radial coordinate $\rho$. The expressions for the
quantities on the left-hand sides of equations (\ref{22}) in metric
(\ref{metrik2}) were obtained in \cite{4}:
$$
W^0_0-W^1_1=\frac{D(\rho D)''''}{3\rho C^4} \, ,
$$
$$
W_{11}=\frac{1}{3C^2D}\left(\frac{D'D'''}{2}-\frac{D''^2}{4}-\frac{DD'''-D'D''}{\rho}-
\frac{DD''+D'^2}{\rho^2}+\frac{2DD'}{\rho^3}-\frac{D^2}{\rho^4}+\frac{1}{\rho^4}\right)
\, .
$$
We obtain the following expressions for the right-hand sides of
equations (\ref{22}) and the scalar curvature $R$:
$$
\begin{array}{l}
\dc {\cal T}^0_0-{\cal
T}^1_1=-\varepsilon+\frac{S_0^2D}{3C}\left(\frac{C'}{C^2}\right)' \, ,
\medskip \\
\dc {\cal T}_{11}=-\frac{S_0^2}{12}\left(\frac{D''}{2D}+\frac{1-D}{\rho^2D}+3F''+\frac{F'D'}{D}
-\frac{2F'}{\rho}-3F'^2\right)+\frac{C^2}{4D}\varepsilon \, , \medskip \\
\dc R=\frac{6(\rho^2DC')'-C(\rho^2(1-D))''}{\rho^2C^3} \, ,
\end{array}
$$
where $F=\log C$. Linearizing equations (\ref{22}) and relation (\ref{1}) in
$c(\rho)$ and $d(\rho)$, we arrive at the following equations:
\begin{equation}\label{2}
-(\rho d)''''=6p\rho c''-\frac{3\rho\varepsilon}{4\alpha_g},
\end{equation}
\begin{equation}\label{3}
\frac{1}{3\rho}\left(d'''+\frac{d''}{\rho}-\frac{2d'}{\rho^2}+\frac{2d}{\rho^3}\right)=
\frac{p}{2}\left(\frac{d''}{2}-\frac{d}{\rho^2}-3c''+\frac{2c'}{\rho}\right)+
\frac{\varepsilon}{16\alpha_g},
\end{equation}
\begin{equation}\label{4}
6(\rho^2c')'=(\rho^2d)''+24q\rho^2+\frac{6}{S_0^2}\varepsilon\rho^2,
\end{equation}
where we have made the notation
$$
p=\frac{S_0^2}{24\alpha_g},\quad q=\lambda S_0^4.
$$

Before solving equations (\ref{2})--(\ref{4}), we should make one reservation.
It is known that, in general relativity, the dynamics of the matter fields and
metric field are closely coupled: the energy-momentum tensor of matter fields
determines the metric while the metric influences the dynamics of matter. As a
result, generally speaking, we cannot arbitrarily set a distribution of matter
$\varepsilon(r)$, $p(r)$ and its dynamics, described by the four-velocity
$u^{\mu}$. This property of cross-influence of the metric on the dynamics of
matter and vice versa is caused by the nonlinearity of the field equations
(and, in particular, by the lack of superposition principle). In the linear
approximation, the metric entering the tensor $T_{\mu\nu}$ is replaced by the
Minkowski metric, and the dynamics of matter is considered against a flat
space-time background and can be set arbitrarily, satisfying only the
conservation laws.\footnote{In a curved space-time, the expression
$\nabla_{\mu}T^{\mu\nu}=0$ cannot be presented in the form
$\partial_{\mu}T^{\mu\nu}=0$ that expresses the conservation laws, and
contains, besides the last ones, the equations of motion for matter. In a flat
space-time, $\nabla_{\mu}T^{\mu\nu}=\partial_{\mu}T^{\mu\nu}$,  and, therefore,
the equations $\nabla_{\mu}T^{\mu\nu}=0$ express only the conservation laws.}
For example, it is easy to verify that Einstein's equations in the linear
approximation allow stationary solutions for any spherically symmetric
distribution of matter at zero pressure. The same is true in the conformal
theory, which follows from the self-consistency of equations (\ref{eq}) in the
linear approximation at $p=0$ and arbitrary $\varepsilon(r)$. In particular,
the system of three equations (\ref{2})--(\ref{4}) for two unknown variables
$c(\rho)$ and $d(\rho)$ is consistent with any dependence $\varepsilon(\rho)$.

Following \cite{3}, we set
\begin{equation}\label{111}
d(\rho)=-2q\rho^2+v(\rho)
\end{equation}
[it allows to exclude the cosmological term $q\rho^2$ from equations
(\ref{2})--(\ref{4})] and pass to new independent functions $\tilde{a}(\rho)$
and $\tilde{b}(\rho)$ which are expressed via $v(\rho)$ and $c(\rho)$ by the
relations [see equations (\ref{55}) and (\ref{111})]:
$$
\tilde{a}(\rho)=v(\rho)-2\rho c'(\rho),\quad
\tilde{b}(\rho)=v(\rho)-2c(\rho).
$$
As a result, equations (\ref{2})--(\ref{4}) take the form
\begin{equation}\label{5}
-\left(\frac{(\rho y)''}{\rho}\right)'=-\frac{18\rho p}{S_0^2}\varepsilon+3p(\tilde{b}-\tilde{a})',
\end{equation}
\begin{equation}\label{6}
\left(\frac{(\rho y)'}{\rho^2}\right)'=\frac{3p}{2}\left(\frac32\rho
\tilde{b}''-y'-\frac{\tilde{a}}
{\rho}\right)+\frac{9p}{2S_0^2}\rho\varepsilon,
\end{equation}
\begin{equation}\label{7}
\left[ \rho \left( \rho\tilde{b}'+2\tilde{a} \right)
\right]'=-\frac{6\rho^2\varepsilon}{S_0^2},
\end{equation}
where $y=y(\rho)=\rho \tilde{b}'(\rho)-\tilde{a}(\rho)$. Integrating equations
(\ref{5}) and (\ref{7}) from $0$ to $\rho$, we obtain
$$
-(\rho y)''=3p\rho(\tilde{b}-\tilde{a})+6p\rho g(\rho)+C_1\rho,
$$
\begin{equation}\label{8}
\rho \tilde{b}'+2\tilde{a}=\varphi(\rho)+C_2/\rho,
\end{equation}
where
$$
g(\rho)=-\frac{3}{S_0^2}\int\limits_0^\rho r\varepsilon(r)\,dr, \qq
\varphi(\rho)=-\frac{3M(\rho)}{2\pi S_0^2\rho},\q
M(\rho)=4\pi\int\limits_0^{\rho}\!\varepsilon(r)r^2dr.
$$
It is necessary to set the constant of integration $C_2$ to zero in (\ref{8})
because, for a non-singular distribution $\varepsilon(\rho)$, the metric should
also be non-singular. Thus equation (\ref{8}) reads
\begin{equation}\label{9}
\rho \tilde{b}'+2\tilde{a}=\varphi(\rho).
\end{equation}
To eliminate $\tilde{a}(\rho)$ from this equation, we divide (\ref{9}) by
$\rho$ and integrate it from $\rho$ to infinity, and then express the integral
of the quantity $\tilde{a}(\rho)/\rho$ from equation (\ref{6}). As a result, we
obtain
\begin{equation}\label{10}
\tilde{b}''+\frac{2}{\rho}\tilde{b}'+p\tilde{b}=-\frac{2\varepsilon}{S_0^2}-\frac{3p}{2\pi
S_0^2}\int\limits_\rho^\infty\!\frac{M(x)}{x^2}\,dx
\end{equation}
(assuming that the function $\tilde{b}(\rho)$ becomes zero at infinity).

The obtained equation can be simply integrated. For this purpose, first, we
find a solution of equation (\ref{10}) for the case of a point source with
density $\varepsilon(\vec \rho)=m\delta(\vec \rho)$. The corresponding metric
function is denoted by $\beta(\rho)$. The equation for $\beta(\rho)$ reads
$$
\Delta\beta(\rho)+p\beta(\rho)={}-\frac{2m}{S_0^2}\delta(\vec \rho)-\frac{3pm}{2\pi
S_0^2\rho}.
$$
To solve this equation, we separate a singular part in $\beta(\rho)$, caused by
the source:
$$
\beta(\rho)=\frac{\alpha}{\rho}+\beta_1(\rho),
$$
where the function $\beta_1(\rho)$ is regular at $\rho=0$. We use
the known relation
$$
\Delta\frac{1}{r}={}-4\pi\delta(\vec r) \, ,
$$
which implies that the constant  $\alpha$ should be set equal to
${m}/{2\pi S_0^2}$. The equation for $\beta_1$
$$
\beta''_1+\frac{2}{\rho}\beta'_1+p \beta_1=-\frac{2pm}{\pi S_0^2\rho}
$$
is easily integrated by the substitution $\beta_1(\rho)=u(\rho)/\rho$ with the
boundary condition $u(0)=0$. Finally, we get
$$
\beta(\rho)=\frac{m}{2\pi S_0^2\rho}+\frac{n}{4\pi S_0^2}\frac{\sin(k\rho)}{\rho}-\frac{2m}{\pi
S_0^2}\frac{1-\cos(k\rho)}{\rho},
$$
where $n=\const$, and $k=\sqrt{p}$ (we consider the case $\alpha_g>0$). For the
metric of a point source, which we call a nucleon, we obtain
\begin{equation}\label{11}
\begin{array}{l}
\dc g_{00}=1-\frac{m}{2\pi S_0^2r}-\frac{n}{4\pi
S_0^2}\frac{\sin(kr)}{r}+\frac{2m}{\pi S_0^2}\frac{1-\cos(kr)}{r}+2qr^2, \medskip \\
\dc g_{11}=1 - \frac{m}{2\pi S_0^2r} + \frac{n}{8\pi S_0^2}
\left[\frac{\sin(kr)}{r}-k\cos(kr)\right] + \frac{m}{\pi S_0^2} \left[k\sin(kr)
- \frac{1-\cos(kr)}{r} \right]\!-2qr^2.
\end{array}
\end{equation}
In what follows, we drop the cosmological term  $qr^2$,  which is known to be
small on the Solar-system scale. The solution (\ref{11}) depends not only on
the nucleon mass $m$, but also on the constant $n$, the physical meaning of
which will not be discussed in this paper.

Now we proceed to the issue of integration of equation (\ref{10}). The solution
$\beta(\rho)$ obtained for a point source plays a role of the Green function
for equation (\ref{10}) in the sense that
\begin{equation}\label{12}
\tilde{b}(\vec r)=\int\beta(\vec r-\vec r\,')n(\vec r\,')\,dV',
\end{equation}
where $n(\vec r)=\varepsilon(\vec r)/m$ is the distribution of nucleons. Due to
spherical symmetry of the matter distribution $n(r)$, we can integrate
(\ref{12}) over the angular variables, which leads to the following expression
for the metric of the distributed source:
\begin{eqnarray}\label{13}
b(r)= - \frac{3}{2\pi S_0^2}\int\limits_r^\infty\frac{M(r')dr'}{r'^{\,2}}+
\frac{8}{S_0^2}\int\limits_r^\infty\!\varepsilon(r')\frac{\sin[k(r-r')]}{kr}r'dr'
\nonumber \\
 + \left[\frac{\eta}{S_0^2}\frac{\sin kr}{kr}+\frac{8}{S_0^2}\frac{\cos
kr}{kr}\right]\int\limits_0^\infty\!\varepsilon(r')\sin(kr')r'dr',
\end{eqnarray}
\begin{eqnarray}\label{14}
a(r) = - \frac{3}{2\pi
S_0^2}\frac{M(r)}{r}+\frac{4}{S_0^2}\int\limits_r^\infty\!\!\varepsilon(r')\!
\left(\frac{\sin[k(r-r')]}{kr}-\cos[k(r-r')]\!\right) \nonumber \\
- \frac{1}{2}\left[\frac{\eta}{S_0^2}\Bigl(\cos kr-\frac{\sin
kr}{kr}\Bigr)-\frac{8}{S_0^2}\Bigl(\sin kr+\frac{\cos
kr}{kr}\Bigr)\right]\int\limits_0^\infty\!\varepsilon(r')\sin(kr')r'dr',
\end{eqnarray}
where $\eta=n/m$ is the dimensionless constant which characterizes a nucleon.
In particular, for a spherically symmetric object (which we call a star) of
mass $M$ and radius $R$ (i.e., $\varepsilon(r>R)=0$), expressions  (\ref{13})
and (\ref{14}) have the form
\begin{equation}\label{15}
\begin{array}{l}
 \dc b(r>R)=-\frac{3M}{2\pi S_0^2r}+\frac{\eta\sin kr+8\cos kr}{krS_0^2}\,C, \medskip \\
 \dc a(r>R)=-\frac{3M}{2\pi S_0^2r}-\frac{C}{2S_0^2}
 \left[\eta\Bigl(\cos kr-\frac{\sin kr}{kr}\Bigr)-8\Bigl(\sin kr+\frac{\cos
kr}{kr}\Bigr)\right],
\end{array}
\end{equation}
where $$C=\int\limits_0^R\varepsilon(r)\sin(kr)rdr.$$

The solution (\ref{15}) improves the exterior solution for the spherically
symmetric source obtained earlier in \cite{3}
\begin{equation}\label{16}
\begin{array}{l}
\dc a(r>R)=\frac{2m}{r}+N\left[\frac{\sin
(kr+\phi)}{r}-k\cos(kr+\phi)\right], \medskip \\
\dc b(r>R)=\frac{2m}{r}+2N\,\frac{\sin (kr+\phi)}{r}.
\end{array}
\end{equation}
It contains three constants of integration $m$, $N$ and $\phi$, whose relation
with the source mass was not found in \cite{3}. Comparing solution (\ref{15})
with (\ref{16}), we arrive at the following relations:
\begin{equation}\label{20}
m=-\frac{3M}{4\pi S_0^2},\quad\phi={\rm arctg\,}(8/\eta),\quad N=\frac{C}{kS_0^2}\left(1+
\frac{\eta^2}{64}\right)^{1/2}.
\end{equation}

\section{Analysis}
Let us analyze the obtained solution (\ref{15}) for the cases $kR\ll 1$, $kR\gg
1$ and $kR\simeq 1$.
\bigskip

\noindent{\bf Case 1:} $\mathbf{kR\ll 1}$.\q As it is known, in the case of a
weak gravitational field, the function $b(\vec r)$ is related to the Newtonian
potential $\Phi(\vec r)$ by the relation $\Phi(\vec r)=-b(\vec r)/2$.
Substituting this expression into (\ref{10}) and taking into account that the
summands $p\tilde{b}$ and $\frac{3p}{2\pi
S_0^2}\int\limits_\rho^\infty\frac{M(x)}{x^2}\,dx$ are small on the scale
$kr\ll 1$, we arrive at a conclusion that equation (\ref{10}) agrees with the
Newton law of gravitation $\Delta \Phi(\vec r)=4\pi G\varepsilon(\vec r)$ if
one sets\footnote{We remind the reader that the function $b(r)$ differs from
$\tilde{b}(r)$ only by the cosmological term $2qr^2$.}
\begin{equation}\label{18}
G=\frac{1}{4\pi S_0^2}.
\end{equation}

Next, in the case under consideration, we can replace $\sin kr$ by $kr$ in the
integrand for the constant $C$, which leads to the value $C=kM/4\pi$. As a
result, solution (\ref{15}) takes the form
\begin{equation}\label{17}
\begin{array}{l}
\dc b(r>R)=\frac{2GM}{r}+{MG\eta}\,\frac{\sin(kr)}{r}-8MG\,\frac{1-\cos(kr)}{r},
\medskip \\
\dc
a(r>R)=-\frac{2GM}{r}\!+\!\frac{MG\eta}{2}\left[\frac{\sin(kr)}{r}-k\cos(kr)\right]\!+
4MG\left[k\sin(kr)\!-\!\frac{1-\cos(kr)}{r}\right]
\end{array}
\end{equation}
and coincides with the solution (\ref{11}) obtained above for a nucleon with
the replacement $m\rightarrow Nm$ and $n\rightarrow Nn$, where $N=M/m$ is the
number of nucleons. Near the source, we have $kr\ll 1$, and the main
contribution to the metric comes from the first summands on the right-hand
sides of solution (\ref{17}), so we can write
$$
b(r>R)=\frac{2GM}{r}\,,\q a(r>R)={}-\frac{2GM}{r}\,,\q kr\ll 1.
$$

The expression for $b(r)$ coincides with the Schwarzschild solution for a
source of mass  $M$ and therefore correctly describes the motion of bodies with
nonrelativistic velocities and gives correct values for the gravitational
redshift of spectral lines. However, the expression for $a(r)$ has wrong sign.
Thus, at small distances from the source, the metric is conformally flat,
which, in particular, leads to a wrong law of the relativistic deflection of
light. Specifically, metric (\ref{17}) leads to the following expression for
the deflection angle $\triangle\varphi$:
$$
\triangle\varphi=\rho\int\limits_\rho^\infty\frac{a(r)-rb'(r)}{r(r^2-\rho^2)^{1/2}}dr= 3\pi r_{g}k
\left[k\rho\left(\frac{\eta}{16}-\frac{1}{\pi}(\ln\frac{k\rho}{2}+\gamma+1/2)\right)+O(k^2\rho^2)\right],
$$
where $\rho$ is the impact parameter, $\gamma$ is the Euler constant, and
$r_{g}=2MG$. This expression essentially differs from the known law\
$\triangle\varphi=2r_g/\rho$. Their ratio is $\sim k^2\rho^2\ln(1/k\rho)\ll 1$.
\medskip

\noindent{\bf Case 2:} $\mathbf{kR\gg 1}$.\q Integrating the expression for the
constant $C$ by parts and taking into account that $\varepsilon(R)$ and its
derivatives become zero at the surface of a star, we obtain
$$
C\simeq{}-\frac{2\varepsilon'(0)}{k^3}\sim{}-kM\left(\frac{1}{kR}\right)^4.
$$
Thus, everywhere outside the star, the ratio of the last summand to the first
one in the expression (\ref{15}) for $b(r)$ is of the order $(1/kR)^4$, and, in
the case $kR\gg 1$, can be neglected. As a result, we have
$$
b(r>R)={}-\frac{3M}{2\pi S_0^2r},\q kR\gg1 \, ,
$$
which corresponds to gravitational repulsion rather than attraction.
\bigskip

\noindent{\bf Case 3:} $\mathbf{kR\simeq 1}$.\q In this case, all summands in
the expression (\ref{15})  for $b(r)$ are of the same order, and the functions
$\sin kr$ and $\cos kr$ essentially vary on the Solar-system scale.
Consequently, we would not have the usual Keplerian planetary orbits, and
strongly elongated orbits of comets would not be closed and would have the
significant perihelion shift per revolution.

Thus, we must conclude that, for all possible values of the
parameter $kR$, the consequences of the conformal gravitation do not
agree with the observational data on the Solar-system scale. This
reasserts the results obtained by Flanagan \cite{2} which were
subjected to criticism by Mannheim \cite{10}.

Above, we considered the case $\alpha_g>0$. In the case $\alpha_g<0$, the
hyperbolic functions  $\sinh kr$ and $\sinh kr$ appear in the solution instead
of $\sin kr$ and $\cos kr$, implying that the metric exponentially blows at
infinity, which is physically forbidden. Although the linear approach is not
valid at large distances, the presence of the growing exponents would imply
that any arbitrarily small mass would create a very strong gravitational field
at distances of the order $r\geq1/\sqrt{-p}$.

\section{Summary}
The results of our investigation are in complete agreement with the conclusions
made in \cite{2}: the conformal theory of gravitation is inconsistent with the
observational data on the Solar-system scale.

\section*{\large Acknowledgements}
We thank Yu. V. Shtanov for helpful conversations.


\begin{thebibliography}{}
\bibitem{7} P.~D.~Mannheim, Ap. J. {\bf 391}, 429 (1992).
\bibitem{1} J.~Wood and W.~Moreau, gr-qc/0102056.
\bibitem{2} \'Eanna \'E.~Flanagan, Phys. Rev. D {\bf 74}, 023002 (2006).
\bibitem{10}P.~D.~Mannheim, Phys. Rev. D {\bf 75}, 124006 (2007).
\bibitem{3} O.~V.~Barabash and Yu.~V.~Shtanov, Phys. Rev. D {\bf60},
064008 (1999).
\bibitem{4}  D.~Kazanas and P.~D.~Mannheim, Ap. J.
 {\bf342}, 635 (1989).
\bibitem{5} R.~Bach, Math. Zeit. {\bf 9}, 110 (1921).
\bibitem{6} J.~D.~Anderson {\it et al.}, Phys. Rev. Lett. {\bf 81}, 2858 (1998).
\bibitem{8} P.~D.~Mannheim, Phys. Rev. D {\bf 58}, 103511 (1998).
\bibitem{9} P.~D.~Mannheim, Gen. Rel. Grav. {\bf 25}, 697 (1993).
\end{thebibliography}
\end{document}